\documentclass[fleqn,usenatbib,useAMS]{mnras}
\usepackage{graphicx}   % Including figure files
\usepackage{amsmath}    % Advanced maths commands
\usepackage{multicol}        % Multi-column entries in tables
\usepackage{bm}     % Bold maths symbols, including upright Greek
\usepackage{pdflscape}  % Landscape pages

\usepackage[T1]{fontenc}
\usepackage{ae,aecompl}

\usepackage{newtxtext,newtxmath}
\title[Tides, anisotropic collapse  massive star formation]{Modification to the Jeans criterion by external tides: Anisotropic fragmentation and formation of filaments}

\author[Guang-Xing Li]{Guang-Xing Li$^{1}$%
\thanks{Contact e-mail: \href{mailto:gxli@ynu.edu.cn}{gxli@ynu.edu.cn}, \href{mailto:ligx.ngc7293@gmail.com}{ligx.ngc7293@gmail.com}}%
\\
$^{1}$South-Western Institute For Astronomy Research, Yunnan University, Kunming, 650600, China}

\date{Last updated 2020 June 10; in original form 2013 September 5}

\pubyear{2020}

\begin{document}
\label{firstpage}
\pagerange{\pageref{firstpage}--\pageref{lastpage}}
\maketitle

% Abstract of the paper
\begin{abstract}
The Jeans criterion sets the foundation of our understanding of gravitational collapse. Jog studied the fragmentation of gas under external tides and derived a dispersion relation
$$
 l' = l_{\rm Jeans} \frac{1} {(1 + \lambda_0' / 4 \pi G \rho_0)^{1/2}} \;.
$$
She further concludes that the Jeans mass is 
$m_{\rm  incorrect}'=m_{\rm Jeans} ( 1/(1 + \lambda_0' / 4 \pi G \rho_0)^{3/2})$. We clarify that due to the inhomogeneous nature of tides, this characteristic mass is incorrect. Under weak tides, the mass is $m \approx \rho\, l_1 l_2 l_3$, where the modifications to Jeans lengths along all three dimensions need to be considered; when the tide is strong enough, collapse can only occur once 1 or 2 dimensions. In the latter case, tides can stretch the gas, leading to the formation of filaments. 

\end{abstract}

% Select between one and six entries from the list of approved keywords.
% Don't make up new ones.
%%%%%%%%%%%%%%%%%%%%%%%%%%%%%%%%%%%%%%%%%%%%%%%%%%

%%%%%%%%%%%%%%%%% BODY OF PAPER %%%%%%%%%%%%%%%%%%

% The MNRAS class isn't designed to include a table of contents, but for this document one is useful.
% I therefore have to do some kludging to make it work without masses of blank space.
% In the standard picture, the fragmentation 
% occurs in a medium of density $\rho$ and velocity dispersion $\sigma_{\rm v}$.
% The crossing time is $t_{\rm cross} = \lambda / \sigma_{\rm v}$  and the
% free-fall time is $t_{\rm ff } = 1 / \sqrt{G \rho}$. The fragmentation length can be determined by letting 
% $t_{\rm cross} = t_{\rm ff}$, and the resulting length scale 

\section{Introduction: Jeans criterion under external tides }\label{sec1} 
\noindent 
The Jeans criterion \citep{1902RSPTA.199....1J} sets the fundament of our understanding of gravitational collapse. A natural consequence of the long-range nature of gravity is the existence of tides.  
The overall importance of tides inside clouds has been demonstrated by \cite{2024MNRAS.528L..52L}, where 
tides play a dominating role in the collapse of $\gtrsim 80 \%$ of the gas mass and  {$\gtrsim 95 \%$ of the volume. Tides that become an indispensable part of the theory of cloud collapse. 
We discuss the role of tides in the Jeans fragmentation.}

Assuming a density of $\rho(x,y,z)$, the gravitational potential is  determined by $\nabla^2 \phi = 4 \pi G \rho$, the tidal tensor is
\begin{equation}\label{eq:tij}
   T_{ij} = \partial_i \partial_j \phi \;.
\end{equation}

The three eigenvalues of the tidal tensor are $\lambda_1$, $\lambda_2$ and $\lambda_3$  where $\lambda_1 < \lambda_2 < \lambda_3$, and 
$
   \lambda_1 + \lambda_2 + \lambda_3 = 4 \pi G \rho$.
Assuming that the gas has a density of $\rho_{\rm gas}$, its contribution to the tidal tensor is 
\begin{equation}
  T_{ij, \rm gas} =\frac{4}{3} \pi G \rho_{\rm gas}   \, {\rm diag} (1,1,1)  \;. 
 \end{equation}

The effect of external tides is described by $T_{ij,\rm external} = T_{ij} - T_{ij,\rm gas}$, and the eigenvalues of $T_{ij,\rm external}$ is 
\begin{equation}
  \lambda_i'= \lambda_{i, \rm external} = \lambda_i - \frac{4}{3} \pi G \rho_{\rm gas} \;.
\end{equation}

\citet{2013MNRAS.434L..56J} studied the Jeans-like gravitational instabilities under the influence of tides in 1D and found the following dispersion relation 
\begin{equation}\label{eq:dispersion}
   w^2 = k^2 c_{\rm s}^2 - 4  \pi G \rho_0 - \lambda_{\rm i, external} \;,
\end{equation}
where the fragmentation length becomes \footnote{Our Eqs. \ref{eq:dispersion} and \ref{eq:length} looks different from the ones presented in \citet{2013MNRAS.434L..56J} as we have adopted different convention. Our $T_{ij}$ is defined in Eq.\ref{eq:tij} differs from the \citet{2013MNRAS.434L..56J} version by a minus sign. }
\begin{equation}\label{eq:length}
   l' =  l_{\rm Jeans} \frac{1} {(1 + \lambda_{\rm i, external} / 4 \pi G \rho_0)^{1/2}} \;.
\end{equation}
where $\lambda_{\rm i, external}$ is a eigenvalue of the tidal tensor. 
When $\lambda_{\rm external} > -4 \pi g \rho_0 $, tides can increase the fragmentation length and when $\lambda_{\rm external} < -4 \pi G \rho_0 $, fragmentation is suppressed by tides. 

\citet{2013MNRAS.434L..56J} concluded that after accounting for the change of the Jeans length, the Jeans mass needs to be modified. She found  
$m_{\rm  incorrect}'=m_{\rm Jeans} ( 1/(1 + \lambda_0' / 4 \pi G \rho_0)^{3/2})$, where it is wrongly assumed that the modifications to the Jeans length are the same along all three dimensions.  This incorrect formula for the Jeans mass is used in a recent study of the role of tides in the fragmentation of molecular clouds  
\cite{2023MNRAS.524.4614Z}. However, in most of the realistic cases, the tidal fields are inhomogeneous \citep{1979ApJ...231....1W}. To derive the Jeans mass, it is necessary to consider the effects of tides along all three dimensions.

\begin{figure}
   \includegraphics*[width=0.5 \textwidth]{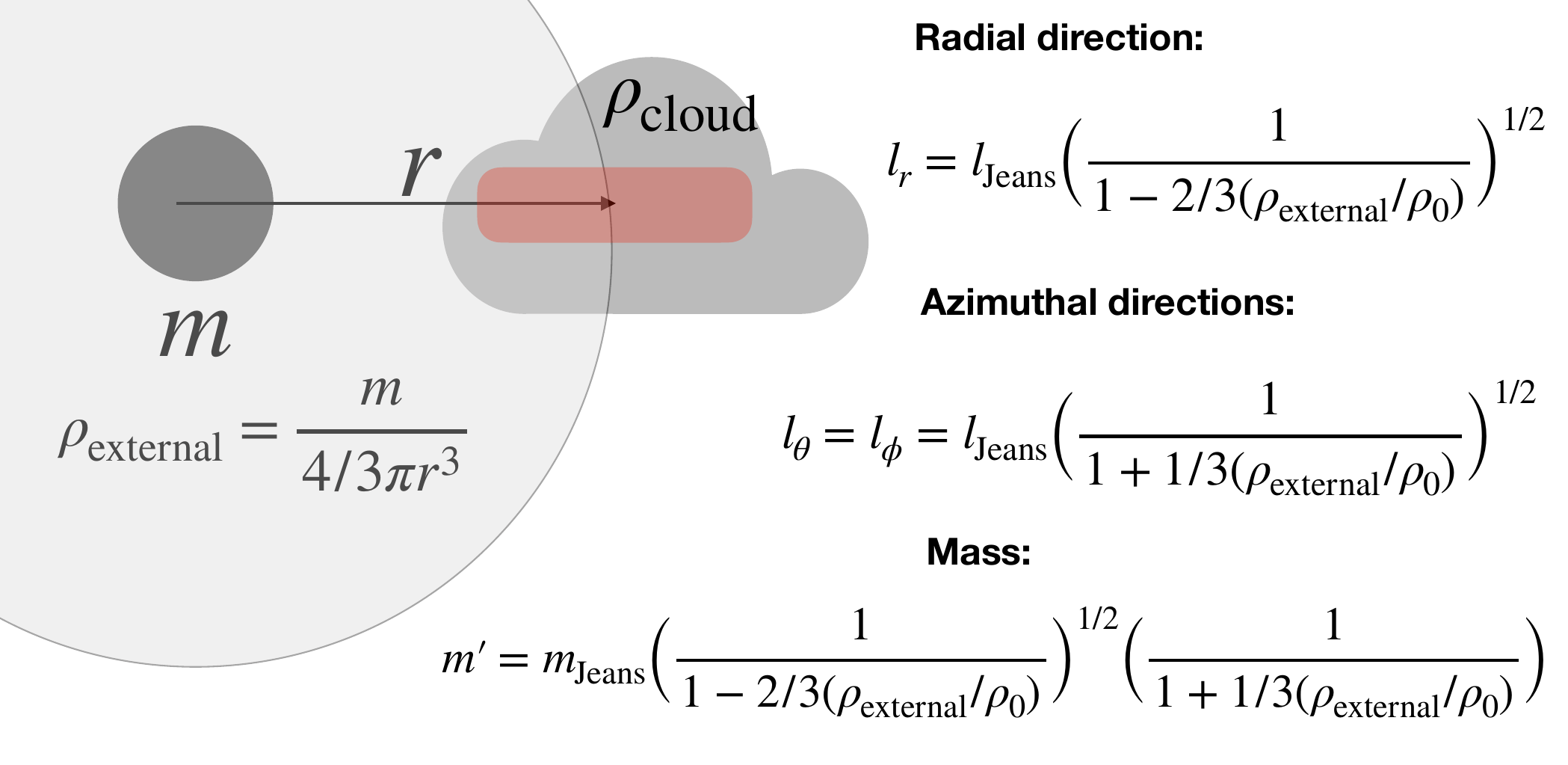}
   \caption{A illustration of the setting we used in Sec. \ref{sec:2}. We consider the effect of tides on a gas path of density $\rho_0$. The source of the tides is an object of mass $m$, and the effect is measured at distance $r$ from the central object. \label{fig:2} }
\end{figure}

\begin{figure*}
   \includegraphics*[width=1 \textwidth]{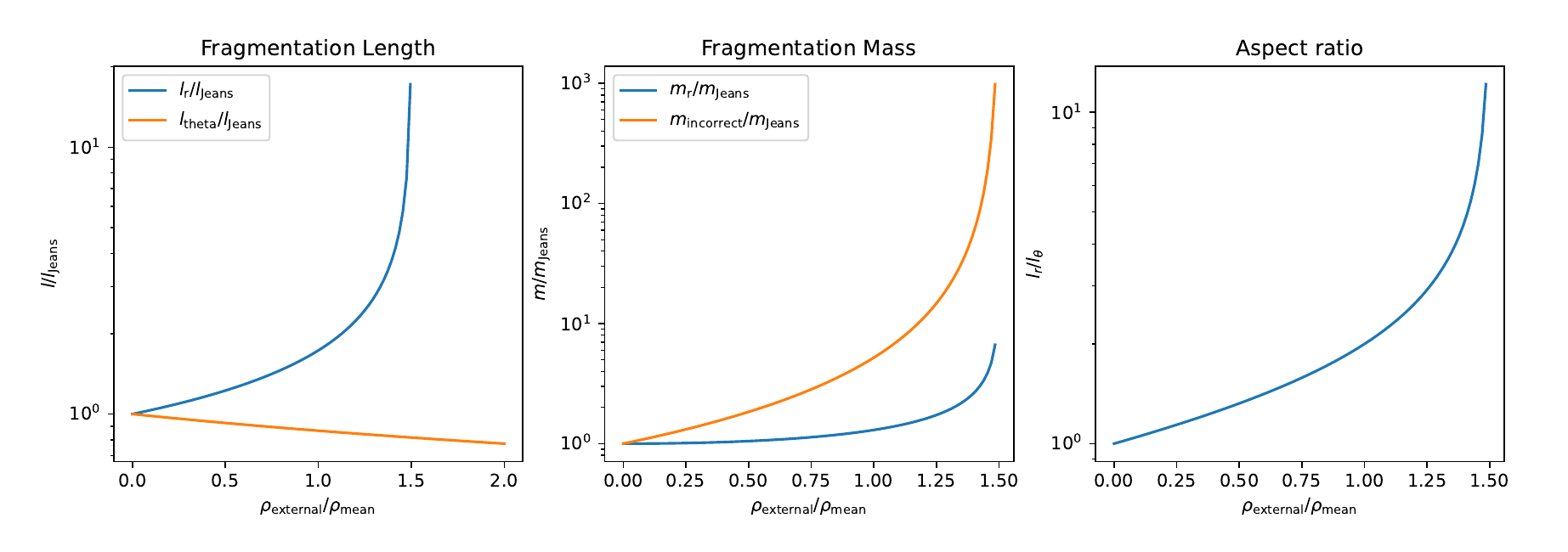}
   \caption{Effect of tides on the Jeans length, Jeans mass, and the aspect ratio of the gravitationally-unstable gas patch. \label{fig:3}}
\end{figure*}

\section{Inhomogeneous collapse and filament formation}\label{sec:2}
For simplicity, we consider the fragmentation of gas of density $\rho$ located at distance $r$ from an object of mass $m$. The setting is illustrated in Fig. \ref{fig:2}.
The external tides are described by the tensor
\begin{equation}
   T_{ij, \rm external} = \frac{4}{3 } \pi G \;{\rm diag}( - 2\rho_{\rm external},\, \rho_{\rm external} ,\, \rho_{\rm external} ) \;,
\end{equation}
where $\rho_{\rm external}  = m / (4/3 \pi r^3)$ is the mean density of the external mass measured at distance $r$. There the first eigenvalues represent the behavior of tides along the radial direction, and the second and their eigenvalues represent the behavior of tides along the azimuthal directions.

\subsection{Inhomogeneous collapse}
When $\rho_0 >   \frac{2}{3}\rho_{\rm external}$, fragmentation is possible along all three directions. 
The shape of the gas patch is $l_r, l_{\theta}, l_{\phi}$, which is elongated along the radial direction.

Along the radial direction, the fragmentation length is
\begin{equation} 
   l_{r} = l_{\rm Jeans} \Big{(} \frac{1}{1 - 2/3 (\rho_{\rm external}/\rho_{0})} \Big{)}^{1/2} \;,
\end{equation}
and along the azimuthal  directions,
\begin{equation}
   l_{\theta} = l_{\phi} = l_{\rm Jeans}  \Big{(}  \frac{1}{1 + 1/3 (\rho_{\rm external} / \rho_0)}  \Big{)}^{1/2}\;.
\end{equation}

The Jeans mass is 
\begin{eqnarray}
   m_{\rm Jeans}' &=& m_{\rm Jeans} f_{\rm tidal} \\ \nonumber
   &=& m_{\rm Jeans}  \Big{(} \frac{1}{1 - 2/3 (\rho_{\rm external}/\rho_{0})} \Big{)}^{1/2} \Big{(}  \frac{1}{1 + 1/3 (\rho_{\rm external} / \rho_0)}  \Big{)} \;.
\end{eqnarray}

The aspect ratio of the gravitationally unstable patch is
\begin{equation}
   A =  \Big{( } \frac{1 + 2/3 (\rho_{\rm external} / \rho_0)}{1 -1/3 (\rho_{\rm external} / \rho_0)}  \Big{)}^{1/2} \;.
\end{equation}

In Fig. \ref{fig:3} we plot the dependence of the fragmentation length, the mass of the fragments, and the aspect ratio of the gravitationally unstable patch.  Contrary to the statement from \cite{2013MNRAS.434L..56J} where the mass can be increased by orders of magnitudes, the tidal effect can only increase the fragmentation mass by a factor of a few. This is because the tide is anisotropic. The increase of the Jeans length along the radial direction is often accompanied by the decrease of the Jean length along the azimuthal direction. The net effect of tides on the Jean mass is therefore much weaker than previously assumed.  

% \citep{1979ApJ...231....1W}.

\subsection{Tidally-driven filamentation}
When $\rho_0 < 2/3 \rho_{\rm external}$, collapse is only possible along two of the azimuthal directions. The collapsing patch has a line mass of
\begin{equation}\label{eq:fila}
   \delta_{\rm ml} = \rho_0 l^2 = \rho_0 l_{\rm Jeans}^2    \frac{1}{1 + 1/3\, (\rho_{\rm external} / \rho_0)} \;, 
\end{equation}
{
where factors such as $\pi$ are neglected for simplicity.} From Eq. \ref{eq:fila}, stronger tides can lead to the formation of finer filaments.

\section{Conclusions}
\citet{2013MNRAS.434L..56J} derived the dispersion relation for gravitational instabilities under the influence of tides. She finds that under the influence of tides, the Jeans length becomes
\begin{equation}
   l' = l_{\rm Jeans} \frac{1} {(1 + \lambda_i' / 4 \pi G \rho_0)^{1/2}}   \;,
\end{equation}
and the Jeans mass becomes 
\begin{equation}
   m'({\rm incorrect}) = m_{\rm Jeans} \frac{1} {(1 + \lambda_i' / 4 \pi G \rho_0)^{3/2}}   \;. 
\end{equation}
The results have been adopted in subsequent research to study the role of tides on the fragmentation of gas clouds \cite{}.

We point out that in most of the realistic cases, the tides are anisotropic. Taking the anisotropy into full account, the Jeans mass is 
\begin{eqnarray}
   m'_{\rm correct} = m_{\rm Jeans} {(1 + \lambda_{1, \rm ext} / 4 \pi G \rho_0)^{-1/2}} \nonumber\\
    {(1 + \lambda_{2, \rm ext} / 4 \pi G \rho_0)^{-1/2}} \nonumber \\
     {(1 + \lambda_{3, \rm ext} / 4 \pi G \rho_0)^{-1/2}}  \;.
\end{eqnarray}
Using a simple spherical model, we demonstrate that the effect of tides can only lead to moderate increases in the characteristic mass. The most noticeable effect of tides is to change the shape of the gravitationally unstable patch into an elongated one, and the further stretch of these elongated patches into filaments.

 \section*{Acknowledgements}
 We thank the referee for very constructive and concise comments.
 GXL acknowledges support from NSFC grant
No. 12273032 and 12033005. 

%%%%%%%%%%%%%%%%%%%%%%%%%%%%%%%%%%%%%%%%%%%%%%%%%%%%%%%
%%% Appendix sections. ??????, ????
%%%%%%%%%%%%%%%%%%%%%%%%%%%%%%%%%%%%%%%%%%%%%%%%%%%%%%%
\section*{Data availability statement}
No proprietary data was used during the preparation of the manuscript.

%\section{Name}

%\end{appendix}

%\begin{appendices}
%\section{Appendix}
%\end{appendices}
%\appendix
\bibliographystyle{mnras}

\bibliography{paper}

%\appendix

% \appendix
% \section{Characteristic mass}\label{sec:appen}

%\end{appendices}

%%%%%%%%%%%%%%%%%%%%%%%%%%%%%%%%%%%%%%%%%%%%%%%%%%

%%%%%%%%%%%%%%%%% APPENDICES %%%%%%%%%%%%%%%%%%%%%

% \appendix
% \section{Journal abbreviations}

%%%%%%%%%%%%%%%%%%%%%%%%%%%%%%%%%%%%%%%%%%%%%%%%%%

% Don't change these lines
\bsp    % typesetting comment
\label{lastpage}
\end{document}